\documentclass[journal=jpcc,manuscript=article]{achemso}
\SectionNumbersOn

\usepackage[version=3]{mhchem} 
\usepackage{hyperref}
\usepackage{tikz}
\usepackage{subcaption}
\usepackage{graphicx}
\usepackage{tabularray}
\usepackage{amssymb}

\usepackage{soul}

\author{Janak Bhandari} 
\author{Ajinkya Shingote} 
\author{Guangming Liu}
\author{Bikram Ghosh}
\author{Hsing-Ta Chen}
\author{Gregory V. Hartland}
\email{ghartlan@nd.edu}

\affiliation[University of Notre Dame]
{Department of Chemistry \& Biochemistry, University of Notre Dame, Notre Dame, IN, 46556, United States}

\title{Energy Transfer and Strong Coupling in Plasmonic Systems}

\date{\today}

\begin{document}

\maketitle

\begin{abstract}
The coupling between plasmon resonances of metal nanostructures and electronic excitations of molecular/semiconductor systems is important in a number of applications, including surface enhanced spectroscopy, hot-electron catalysis, plasmon-induced resonance energy transfer and the creation of hybrid light-matter states. The past several decades has seen a large number of experimental studies of these effects, as well as the development of theory to explain the results. In this Perspectives article we use classical electromagnetic (EM) simulations to examine how the localized surface plasmon resonances (LSPRs) of metal nanoparticles and propagating surface plasmon polaritons (SPPs) in extended metal nanostructures (wires/plates/films) are affected by interactions with a molecular/semiconductor system. The classical EM simulations are complemented by a quantum model for plasmon-exciton coupling based on the macroscopic quantum electrodynamics (MQED) framework, along with Holstein vibrations. The goals are to show how LSPR spectra and SPP dispersion curves/attenuation constants changes as the strength of the molecular/semiconductor electronic transition increases and the system evolves from an energy transfer limit to strong coupling, and to provide an understanding of the decay processes from both classical EM and MQED perspectives.
\end{abstract}

\section{Introduction}

The interaction of light with metal nanostructures can excite coherent motions of the conduction electrons, which are commonly termed ``plasmons"~\cite{kriebig1995optical, maier2007plasmonics, pelton2013introduction}. In metal particles, localized surface plasmon resonances (LSPRs) occur that strongly absorb and scatter light~\cite{kriebig1995optical, maier2007plasmonics, pelton2013introduction}. In extended metal nanostructures, such as nanowires or nanoplates, propagating surface plasmon polaritons (SPPs) can be excited that have a well-defined momentum, but no resonance~\cite{maier2007plasmonics, pelton2013introduction, johns2017dynamics, wei2018plasmon}. LSPRs and SPPs are responsible for many of the useful and unusual properties of metal nanoparticles, such as biomolecule sensing assays and surface-enhanced spectroscopies~\cite{li2017core,wang2020fundamental,hang2024plasmonic}, photothermal therapy~\cite{cui2023photothermal}, and hot electron catalysis~\cite{zhou2024plasmon, hartland2017what}. This has spurred significant physical chemistry research to understand the photo-physics of metal nanostructures, especially how LSPR spectra and SPP dispersion curves depend on the size and shape of the structures, and their environment~\cite{johns2017dynamics, wei2018plasmon, hang2024plasmonic, wang2020fundamental, li2017core, cui2023photothermal, zhou2024plasmon, hartland2017what}.

One of the key issues in metal nanoparticle spectroscopy is plasmon damping. For nanoparticles this is reflected in the width of the LSPR~\cite{sonnichsen2002drastic}. The LSPR becomes broader at large sizes due to radiation damping, and at small sizes due to electron-surface scattering~\cite{kriebig1995optical, novo2006contributions, hu2008dark}. Radiation damping occurs because large particles scatter light very efficiently, so much so that scattering becomes a significant energy loss mechanism. Both these effects are well known: Mie's 1908 paper solving Maxwell's equations for a spherical particle reproduces radiation damping~\cite{bohren1983absorption, mie1908articles}, and electron-surface scattering was observed in the late 1950s/early 1960s~\cite{vonfragstein1958uber, doremus1964optical}, and extensively studied by Kriebig and co-workers for a variety of different systems in the 1970s~\cite{kriebig1995optical, hovel1993width}. Electron-surface scattering can be explained by adding a surface scattering term to the Drude damping component of the dielectric constant of the particle, that is proportional to the surface-to-volume ratio~\cite{kriebig1995optical, vonfragstein1958uber, doremus1964optical, hartland2011optical, coronado2003surface}. This represents a size-dependent change in the material properties of the particle, which predates observations of these effects in semiconductors by several decades.\cite{ekimov1981quantum,rossetti1982electron} Electron-surface scattering, and associated chemical interface damping (CID) effects~\cite{besteiro2019fast, wilson2020light,lee2021chemical, stefancu2024electronic}, are of current interest because of their importance in hot electron catalysis~\cite{mukherjee2013hot, aslam2018catalytic, carlin2023nanoscale, dong2023plasmonic}.

For metal nanostructures in close proximity with molecules or semiconductors that have overlapping resonances, energy exchange with the plasmon can occur, an effect that is known as plasmon-induced resonance energy transfer (PIRET)~\cite{oh2026plasmon,Collines_PIRET_2021_ACSNano,Oh_PIRET_2023_ACSNano,Jones_PIRET_2023_JPCL}. PIRET is important in applications where plasmons are used to amplify absorption or emission in the molecular/semiconductor system~\cite{li2017plasmon}. The rate of energy transfer depends on the strength of the molecular/semiconductor absorption~\cite{oh2026plasmon}. For systems with very strong absorption bands, the rate of energy exchange can exceed the plasmon decay rate for the bare metal~\cite{torma2015strong, baranov2018novel, sun2021strong}. In this case, strongly coupled plasmon-exciton states can be formed, that are mixtures of the molecular/semiconductor excitation and the plasmon~\cite{torma2015strong, baranov2018novel, sun2021strong}. Coupled plasmon-exciton states were first observed for SPPs on Ag films~\cite{pockrand1982exciton, bellessa2004strong}, and in the past decade there have been a number of studies of these states for particles, through both ensemble and single particle measurements~\cite{fofang2008plexcitonic, zengin2013approaching}. In this Perspectives article we examine plasmon-exciton energy exchange in two complementary approaches.  First, classical electromagnetic (EM) simulations are used to calculate LSPR spectra, and SPP dispersion curves/attenuation constants for Au nanostructures coupled to a model dye-doped polymer film. Second, we examine plasmon-exciton coupling through a quantum model grounded in macroscopic quantum electrodynamics (MQED),~\cite{dung_three-dimensional_1998,scheel_spontaneous_1999,scheel_macroscopic_2009,hsu_chemistry_2025}  which can, in principle, generate a microscopic (state-resolved) picture of polariton dynamics.  There are several goals. The first is to provide a simple understanding of energy transfer and plasmon-exciton coupling, especially how these processes affect the absorption and scattering from the system. The second is to examine how the system transitions from the energy transfer limit where the plasmon is damped by interactions with the molecular/semiconductor species, to the creation of well-defined polariton states with an avoided crossing and associated Rabi splitting~\cite{torma2015strong, baranov2018novel, sun2021strong}.    

While the focus of this Perspectives article is on EM simulations and MQED calculations and what can be learned from them, it is important to consider how the experimental measurements are done. For LSPRs information about damping is best obtained by recording spectra of single particles~\cite{crut2014optical, olson2015optical, joplin2018imaging}. This can be achieved by either absorption, extinction or scattering experiments, although the majority of the studies published to date use light scattering~\cite{crut2014optical, olson2015optical, joplin2018imaging}. These experiments are typically performed on fairly large particles, for example, nanorods with widths on the order of 20--40~nm. For SPPs information about the wavevector and the attenuation are needed to characterize the coupling strength and damping.\cite{fukui1979lifetimes,berini2006figures,beane2019light} This is most conveniently done by light scattering measurements of leaky SPP modes (see Figure~4(a) below)~\cite{song2011imaging, wang2014controlling, yang2016identification, zhang2018extending, beane2019light}. Fairly large dimensions are again needed for these experiments, so that the SPP propagation lengths are long enough to be measured in a far--field optical experiment~\cite{johns2017dynamics, wei2018plasmon}. These practical considerations dictate the choice of the dimensions of the nanostructures in the EM simulations.

Finally, before starting, it is also useful to remember the classical EM explanations for why LSPRs and SPPs arise in metal nanostructures. For spherical particles that are much smaller than the wavelength of light, it is straightforward to show that the polarizability is given by $\alpha = 4\pi a^{3}(\epsilon_{1}-\epsilon_{m})/(\epsilon_{1}+2\epsilon_{m})$, where $a$ is the radius, $\epsilon_{1}$ is the dielectric constant for the particle and $\epsilon_{m}$ is the dielectric constant of the surrounding medium~\cite{kriebig1995optical, maier2007plasmonics, pelton2013introduction}. There is a large polarizability and, consequently, large absorption and scattering when $\epsilon_{1} \rightarrow -2\epsilon_{m}$. This is the condition for the LSPR of a sphere. Likewise, SPPs are guided waves at metal-dielectric interfaces with the form $e^{ikz}e^{-y/\delta}$ (propagating along the surface in the $z$ direction, and decaying away from the surface in the $y$ direction). These waves exist when $\delta_{1}/\delta_{m} = -\epsilon_{m}/\epsilon_{1}$. Because the real component of the dielectric constant of noble metals (Ag, Au and Cu) is negative in the visible through to the infrared, these systems can support both LSPRs and SPPs.  

\section{Classical Electromagnetic Simulations}

There are a variety of methods for calculating the optical properties of metal nanostructures~\cite{jensen1999electrodynamics,shatz2008methods,Myroshnychenko2008modelling,BEM2012}. Here we use finite element simulations implemented through COMSOL Multiphysics. To investigate LSPRs, a three-dimensional, frequency domain simulation is performed to calculate the absorption, scattering and extinction spectra of Au nanorods with different aspect ratios. For most of the calculations the nanorod radius was set to $R_{0} = 15$~nm, and the length $L_{0}$ was varied to give different aspect ratios $\text{AR} = 1 + L_{0}/2R_{0}$ and, therefore, different LSPR frequencies, see Figure~1(a). In our calculations the nanorods are in a homogeneous environment, with a uniform dye-doped polymer coating, and are excited by an $(\hat{x}/\hat{y})$ polarized electric field that propagates in the $z$-direction: $\vec{E} = E_{0} e^{-i \cdot k_{0} \cdot n_{\text{med}} \cdot z} (\hat{x}+\hat{y})/{\sqrt{2}}$
where $n_{\text{med}}$ is the refractive index of the surrounding medium (taken to be water) and $k_{0} = 2\pi/\lambda$. The absorption cross-sections $\sigma_{\text{abs}}$ were calculated by integrating the resistive heating induced by the electric field ($Q_{\text{rh}}$ in the language of COMSOL) over the Au and polymer domains, and dividing by the power density of the electric field. For the scattering cross-sections $\sigma_{\text{scat}}$, the scattered power is integrated over a sphere that surrounds the particle. The extinction cross-section $\sigma_{\text{ext}}$ is then obtained from $\sigma_{\text{ext}} = \sigma_{\text{abs}} + \sigma_{\text{scat}}$. Spectra are calculated by performing a "parametric sweep" over different wavelengths. For some simulations a spacer layer was included between the Au and the dye-doped polymer layer.

\begin{figure}[htbp]
\centering
\includegraphics[width=\linewidth]{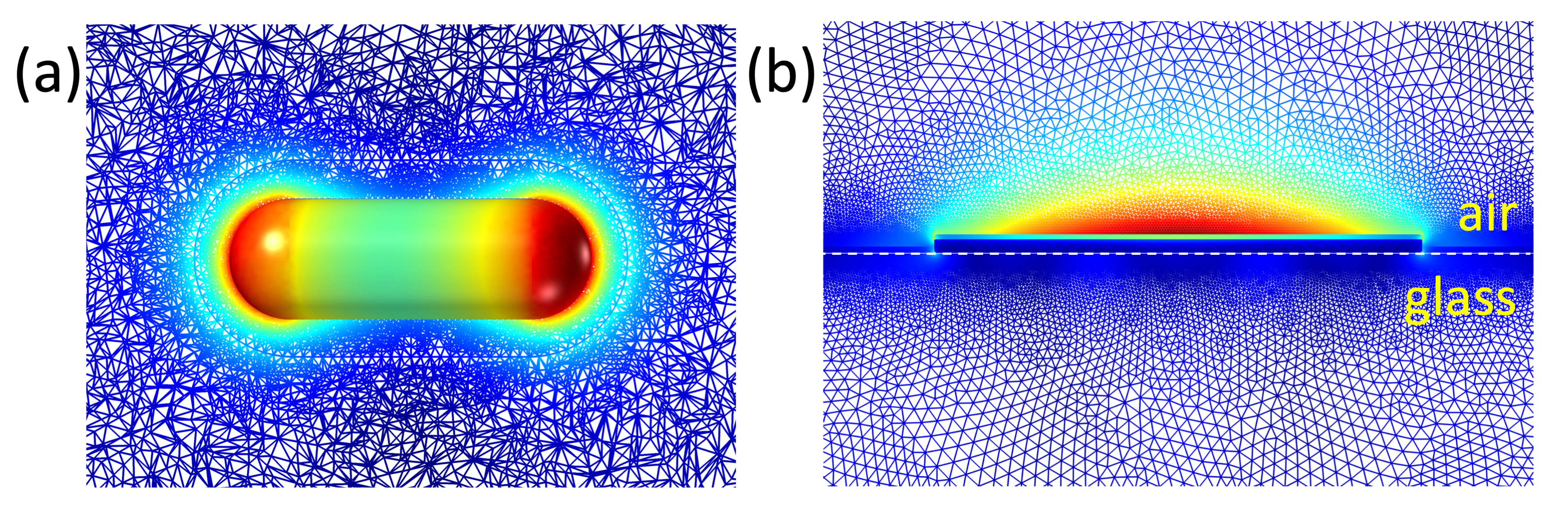}
\caption{Model systems for EM simulations of energy transfer between plasmons and molecules/semiconductors. (a) Three-dimensional Au nanorod in a homogeneous environment. The nanorod is modeled as a cylinder (length $L_{0}$ and radius $R_{0}$) with spherical end caps. The dashed line shows the outline of the dye-doped polymer shell.   (b) Two-dimensional Au nanostripe on a glass surface. The nanostripe is 50 nm thick with a 20 nm polymer layer at the Au-air interface. The plots show cartoons of the mesh used for the finite element simulations, and the color represents the magnitude of the electromagnetic field.  }
\label{fig1}
\end{figure}

To investigate SPPs, mode analysis calculations were performed in a two-dimensional frequency domain simulation in COMSOL~\cite{johns2017dynamics, wei2018plasmon}. These calculations yield the complex effective index for guided modes in the structure $n_{\text{eff}} - i \alpha_{\text{SPP}}/k_{0}$. The real part gives the SPP wavevector $k_{\text{SPP}} = n_{\text{eff}}k_{0}$ and the imaginary part gives the total attenuation constant $\alpha_{\text{SPP}}$, which is related to the SPP propagation length by $L_{\text{SPP}} = 1/(2\alpha_{\text{SPP}})$~\cite{wei2018plasmon, beane2019light, bhandari2025energy, ghosh2025energy}. In this contribution the SPP damping is discussed in terms of attenuation, rather than propagation lengths or lifetimes~\cite{beane2019light, bhandari2025energy, ghosh2025energy, bhandari2026plasmon}, as the attenuation can be broken up into contributions from scattering or absorption in the Au/polymer domains from the relative powers dissipated by the different processes. Specifically, the power dissipated by radiation was calculated from the line integral of the time-averaged Poynting vector $\langle\vec{S}\rangle$ over a circle that encloses the nanostripe: $\oint \vec{n} \cdot \langle\vec{S}\rangle \, dl$ where $\vec{n}$ is the outward normal unit vector. The powers dissipated by resistive heating (optical absorption) in the metal or dye-polymer layers were calculated by $\int Q_{\text{rh}} \, dA$ where $Q_{\text{rh}}$ is the resistive losses in the material and the integral is over the metal or dye-polymer domains~\cite{bhandari2025energy, ghosh2025energy, bhandari2026plasmon}. The different contributions to the SPP attenuation were then calculated by $\alpha_{i} = \eta_{i} \alpha_{\text{SPP}}$ where $\eta_{i}$ are the relative power losses for radiation damping ($\eta_{\text{rad}}$) and resistive heating ($\eta_{\text{Au}}$ and $\eta_{\text{dye}}$)~\cite{bhandari2025energy, ghosh2025energy, bhandari2026plasmon}. Note that the two-dimensional geometry in these simulations strictly corresponds to infinitely long structures, which is appropriate when the dimensions are much greater than the SPP propagation lengths. The specific system modeled was a ``nanostripe'' structure on a glass substrate, see Figure~1(b)~\cite{beane2019light, bhandari2025energy, ghosh2025energy, brown2020chemical}. The dimensions were chosen to be a few $\mu\text{m}$ wide, with a height of 50~nm. Changing the dimensions of the nanostripes causes changes in both $n_{\text{eff}}$ and $\alpha_{\text{SPP}}$. However, the form of the SPP field is similar, so the physics of the plasmon-molecule interaction does not depend strongly on size. A 3~nm thick titanium wetting layer was included in the simulations to match the experimental systems~\cite{bhandari2025energy, ghosh2025energy, bhandari2026plasmon, devkota2019making}.

The dielectric constant data for Au from Ref.~\citenum{olmon2012optical} was used for the calculations, and a Drude-Lorentz oscillator was used the model the dye layer~\cite{saleh2007fundamentals, pockrand1979exciton}. Specifically, the dye is considered to be mixed in a polymer film, and the dielectric constant of the film $\epsilon_{\text{dye}}(\omega)$ was calculated by:
\begin{equation}
\epsilon_{\text{dye}}(\omega) = \epsilon_{\infty} + \frac{f\omega_{0}^{2}}{\omega_{0}^{2} - \omega^{2} - i\gamma\omega}
\end{equation}
where $\epsilon_{\infty}$ is the background dielectric constant, $\omega_{0}$ and $\gamma$ give the resonance frequency and damping of the transition, and $f$ is the effective oscillator strength. The parameters used in the calculations were $\epsilon_{\infty} = 2.1$, $\omega_{0} = 1.5$~eV and $f = 0.1$. $\gamma$ was varied to give broad or sharp resonances. Plots of $\text{Re}[\epsilon_{\text{dye}}(\omega)]$ and $\text{Im}[\epsilon_{\text{dye}}(\omega)]$, as well as $n(\omega) = \text{Re}[\sqrt{\epsilon_{\text{dye}}(\omega)}]$ and $k(\omega) = \text{Im}[\sqrt{\epsilon_{\text{dye}}(\omega)}]$,\cite{saleh2007fundamentals} which are the parameters needed for COMSOL, are presented in the Supporting Information for $\gamma = 0.03, 0.05, 0.20,$ and $0.40$~eV. As will be shown below, the width of the absorption spectrum of the dye (and consequently the intensity at the peak) is the major factor for determining the properties of the coupled system. Note that even though the Drude-Lorentz oscillator model is more appropriate for molecular systems than semiconductors, the conclusions about energy transfer and strong coupling from our analysis are valid for both systems.

\subsection{Localized Surface Plasmon Resonances}

Figure~2(a) shows the absorption, scattering and extinction spectra of a 15~nm radius Au nanorod with an aspect ratio of 3.33, coated with either a 10~nm polymer layer with no dye, or a dye-polymer layer with a broad resonance ($\gamma = 0.4$~eV). In this calculation the aspect ratio of the nanorod was chosen so that the absorption spectrum of the nanorod matches that of the dye. Coating the nanorod with an absorbing dye-polymer layer damps the LSPR resonance. Fitting the spectra to Lorentzian functions yields the LSPR resonance frequencies $\Omega_{R}$ and linewidths $\Delta\Omega_{R}$. For the structures examined in this paper, the values of $\Omega_{R}$ and $\Delta\Omega_{R}$ are the same for the absorption, extinction and scattering spectra. This is not always true: differences in $\Omega_{R}$ and $\Delta\Omega_{R}$ for absorption/scattering/extinction can occur for larger structures, where retardation effects start to become significant~\cite{west2023nonlinear}. The fact that the absorption, scattering and extinction spectra can be described by the same Lorentzian function also means that it is appropriate to relate $\Delta\Omega_{R}$ to the dephasing time for the plasmon $T_{2} = 2\hbar/\Delta\Omega_{R}$~\cite{sonnichsen2002drastic}.

\begin{figure}[htbp]
\centering
\includegraphics[width=\linewidth]{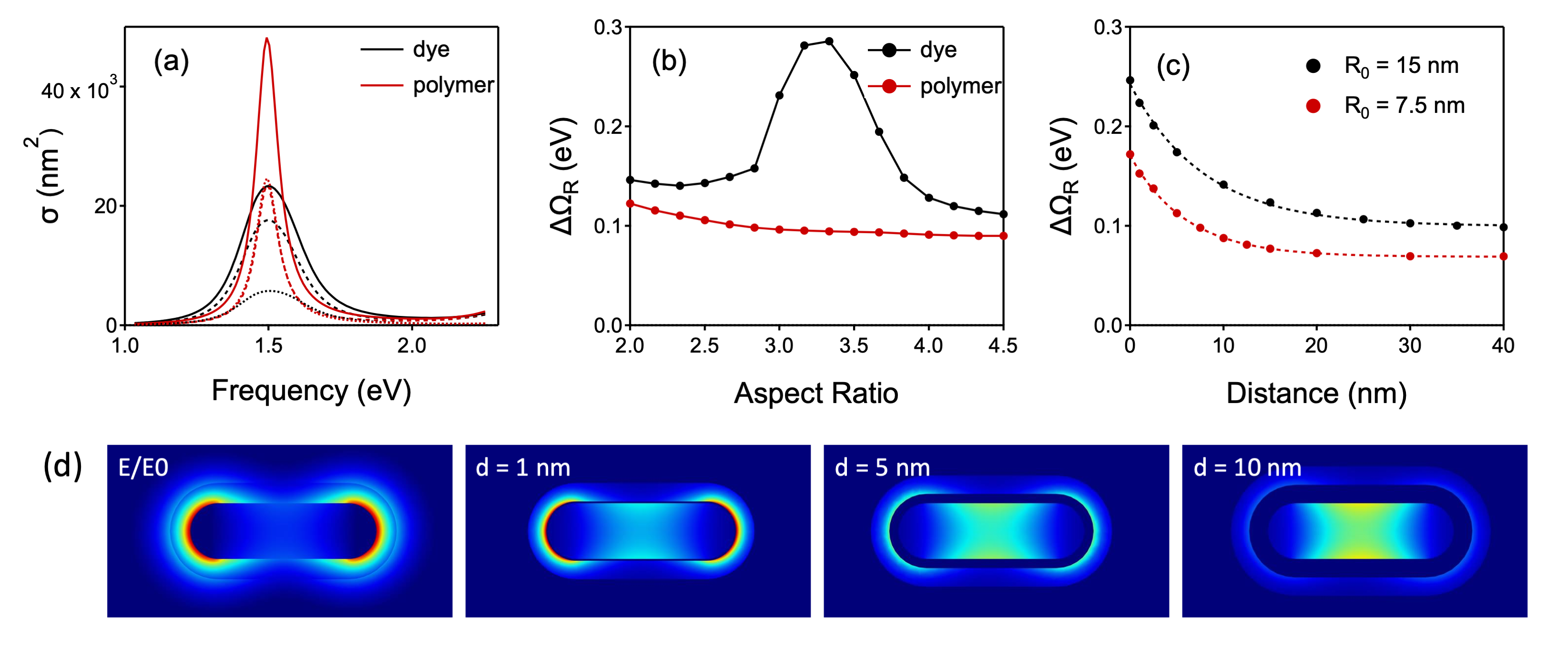}
\caption{(a) Extinction (solid lines), absorption (dashed lines) and scattering spectra (dotted lines) versus frequency for Au nanorods with an aspect ratio of $\text{AR}=3.33$, and a 10~nm thick dye layer with $\gamma=0.4$~eV (``dye''; black lines), or with a polymer shell with no dye (``polymer''; red lines). (b) Full-width-at-half-maximum $\Delta\Omega_{R}$ versus aspect ratio for dye coated and polymer coated Au nanorods with different aspect ratios, same color scheme as panel (a). (c) $\Delta\Omega_{R}$ versus spacer layer thickness for dye coated nanorods ($\text{AR}=3.33$) with $R_0=15$~nm (black markers) or $R_0=7.5$~nm (red markers). The dashed lines are fits using a double exponential.  (d) Left: plot of the normalized electric field $E/E_0$ for a Au nanorod with $\text{AR}=3.33$ enclosed in a 10~nm thick dye layer with $\gamma=0.4$~eV and a 1~nm thick spacer layer. Right: plots of the resistive heating for nanorods with 1~nm, 5~nm and 10~nm thick spacer layers.  The images were generated at an excitation frequency of $\omega=1.5$~eV.}
\label{fig2}
\end{figure}

Figure~2(b) shows a plot of $\Delta\Omega_{R}$ versus aspect ratio for nanorods coated with a dye-polymer layer with $\gamma=0.4$~eV, and polymer coated nanorods with no dye. For the polymer-only coated nanorods, $\Delta\Omega_{R}$ decreases as the aspect ratio increases because the LSPR moves away from the gold interband transitions, which are a strong source of damping~\cite{kriebig1995optical, hartland2011optical}. For the dye-polymer coated nanorods, $\Delta\Omega_{R}$ goes through a maximum at $\text{AR} \approx 3.33$, which is where the nanorod LSPR is resonant with the dye transition. At high and low aspect ratios the effect of the dye layer is reduced: essentially the $\Delta\Omega_{R}$ versus aspect ratio plot traces the absorption spectrum of the dye. The way the plasmon linewidth changes when a spacer layer is included between the Au nanorod and the dye is shown in Figure~2(c). $\Delta\Omega_{R}$ decreases with increasing thickness of the spacer layer, with a length scale that depends on the diameter of the nanorod~\cite{oh2026plasmon}.

These results can be understood in the following way: for this choice of parameters for the dye absorption spectrum, the LSPR is only weakly coupled to the dye resonance. The nanorod acts as an optical antenna with a field that extends over a distance on the order of the nanorod diameter~\cite{hartland2011optical, jensen1999electrodynamics}. Spatial overlap of the LSPR field with the dye layer causes optical absorption in the dye, which creates an additional dephasing process that broadens the LSPR resonance. Figure~2(d) shows plots of the electric field and optical absorption/resistive heating for different thickness spacer layers between the nanorod and the dye layer that demonstrate this effect. The weaker penetration of the LSPR field from the nanorod into the dye layer when the spacer layer is present reduces the amount of absorption due in the dye and, therefore, the amount of LSPR damping induced by the dye.

The results in Figure~2 with a $\gamma = 0.4$~eV damping constant for the Drude-Lorentz oscillator correspond to the energy transfer limit for plasmons coupled to electronic transitions of the molecular layer~\cite{oh2026plasmon}. It is important to note that when we say energy transfer here, we do not mean that the nanoparticle absorbs light and then transfers energy to the dye layer, which is the canonical way of thinking about Förster resonance energy transfer.\cite{FRET_2009_review} Photons are absorbed by the particle-dye system, and the LSPR resonance of the system is broadened by energy dissipation into states associated with the dye layer.

Decreasing the damping parameter for the dye increases the peak intensity of the electronic transition and, thus, the rate of energy exchange between the LSPR and the dye resonance. This can lead to strong coupling and the creation of polariton states~\cite{torma2015strong, baranov2018novel, sun2021strong}. Figures~3(a)--(c) show calculated spectra for nanorods with different aspect ratios for a Drude-Lorentz oscillator with $\gamma = 0.05$~eV, with the other parameters ($f$ and $\omega_{0}$) the same as in Figure~2. In this case the LSPR is split into upper and lower polariton states ($P_{+}$ and $P_{-}$)~\cite{torma2015strong, baranov2018novel, sun2021strong}. At small aspect ratios ($\text{AR} = 2$, Fig.~3(a)) the higher frequency peak shows strong absorption and scattering (a property of plasmonic systems), and can be associated with the nanorod LSPR. The lower frequency peak does not significantly scatter light, and corresponds to the molecular excitation. Note that the absorption cross-section for the lower frequency peak is much larger (by several orders of magnitude) than that of a dye-polymer layer without the nanorod. This increase in absorption is due to antenna effects from the nanorod. Conversely, at large aspect ratios ($\text{AR} = 4.5$, Fig.~3(c)), the lower frequency peak scatters light strongly and, thus, correlates with the nanorod LSPR. The higher frequency peak now corresponds to the molecular excitation. At intermediate aspect ratios ($\text{AR} = 3.33$, Fig.~3(b)), both peaks show strong absorption and scattering contributions, which indicates hybridization of the LSPR with the electronic transition of the dye, and the sharing of properties. Note that a small peak can be seen between the upper and lower polariton peaks at the avoided crossing region (Fig.~3(b)). This is due to dye that is not coupled to the LSPR.

\begin{figure}[htbp]
\centering
\includegraphics[width=\linewidth]{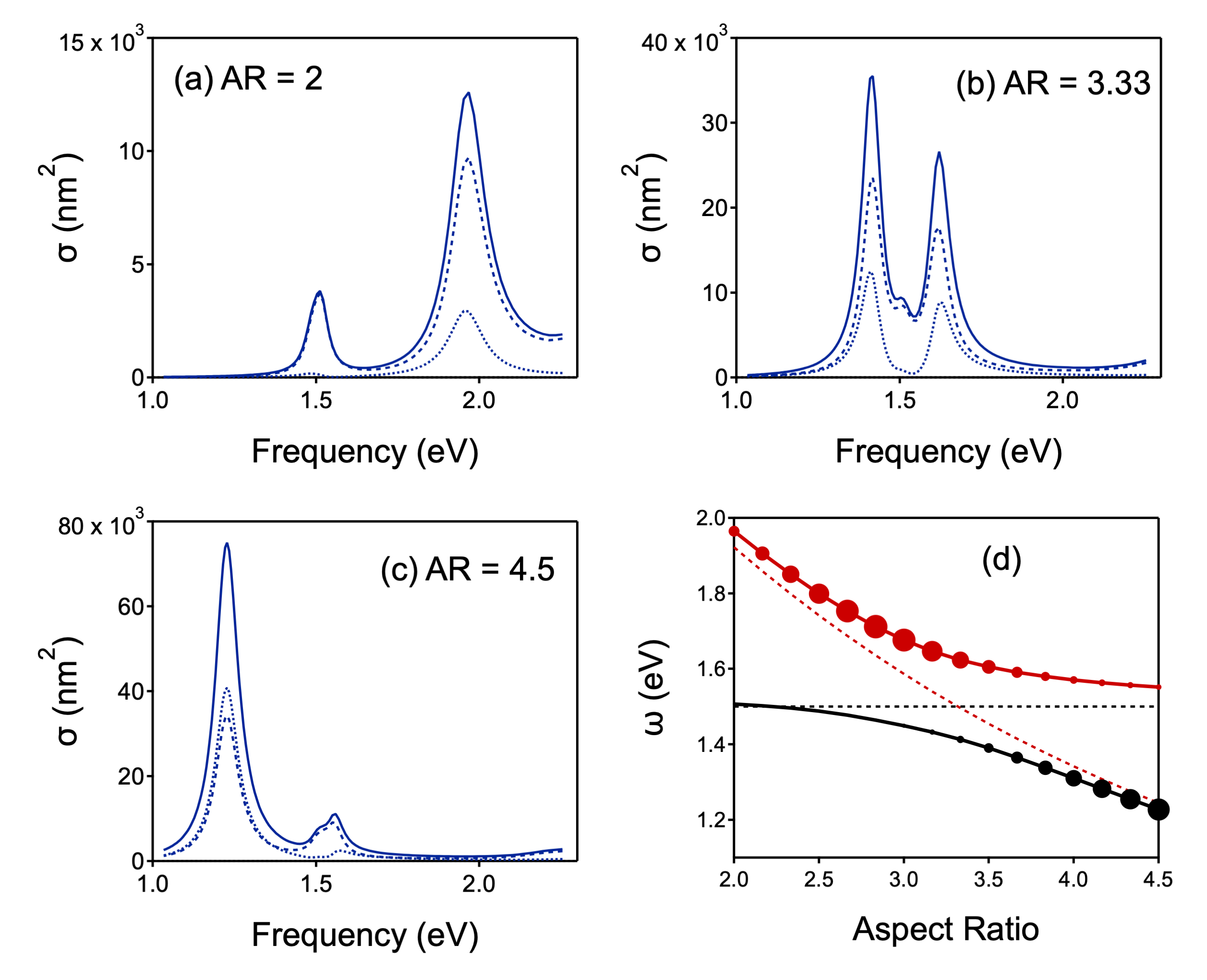}
\caption{(a)--(c) Extinction (solid lines), absorption (dashed lines) and scattering spectra (dotted lines) for Au nanorods with different aspect ratios coupled to a dye-polymer layer with $\gamma=0.05$~eV. Note the different $y-$axis scales for the different panels.  (d) Frequency versus aspect ratio plot. The dashed red and black lines are the resonance frequencies for the uncoupled LSPR and dye resonance, respectively. The size of the symbols for the $P_+$ and $P_-$ branches indicate the relative magnitude of the scattering cross-section for the transition.}
\label{fig3}
\end{figure}

A plot of the frequency of the two transitions as a function of aspect ratio is presented in Fig.~3(d). The size of the symbols for the $P_{+}$ and $P_{-}$ branches in Fig.~3(d) represent the intensity of the scattering contribution to the extinction, which is a proxy for LSPR character. The red and black dashed lines in Fig.~3(d) are the frequencies of the uncoupled states. The system shows an avoided crossing with a Rabi frequency of 210~meV, and the identity of the states (how much they scatter light) switches in the crossing region. Note that the linewidths of the $P_{+}$ and $P_{-}$ resonances $\Delta\Omega_{R}$ are equal in the avoided coupling region, as expected from the simple coupled oscillator model~\cite{torma2015strong,cohen2005quantum}. This reflects that the upper and lower polariton states are equal mixtures of the LSPR and molecular excitation when the two states are strongly coupled~\cite{torma2015strong, baranov2018novel, sun2021strong}. The integrated oscillator strength of the dye is the same for the simulations for Figs.~2 and 3. Whether the system shows energy transfer or polariton formation is controlled by the sharpness of the dye electronic transition in our model, which determines the intensity at the peak of the absorption band.

\subsection{Propagating Surface Plasmon Polaritons}

The propagating SPPs associated with extended metal nanostructures are also affected by the presence of absorbing species at their surface. These systems are typically characterized by measuring frequency versus wavevector dispersion curves, and strong plasmon-exciton coupling effects cause avoided crossings in these curves~\cite{pockrand1982exciton, bellessa2004strong}. This was first observed for Ag films by Pockrand and co-workers using total internal reflection measurements in the ``Kretschmann" geometry (excitation through a prism to match the wavevector of light with the SPP wavevector)~\cite{pockrand1982exciton}. For thin film studies it is hard to quantify the attenuation induced by the molecular layer, which is a motivation for single nanostructure experiments. Specifically, for single metal nanowires SPPs can be launched by focusing a light source at the end of the structure, where the break in symmetry relaxes the momentum matching condition for coupling between SPPs and photons~\cite{dickson2000unidirectional,Sanders_2006_SPP}. Imaging measurements can then give the SPP attenuation constant~\cite{johns2017dynamics, solis2010bleach, wild2012propagation,Yu_2014_SPPs}.

For nanostructures on a surface there are two types of SPP modes:``bound'' modes that propagate at the metal-substrate interface and ``leaky'' modes that propagate at the metal-air interface~\cite{song2011imaging, wang2014controlling, yang2016identification,johns2017dynamics, wei2018plasmon}. Plots of the electric field for the lowest order bound and leaky modes of a $4~\mu\text{m}$ wide gold nanostripe (a model system that is relatively easy to create by nanofabrication) are shown in Figs.~4(a) and (b)~\cite{bhandari2025energy}. Because the field for the leaky mode is localized at the top of the structure, it has a much stronger interaction with adsorbed species~\cite{bhandari2025energy}. The wavevector and attenuation constant for the leaky mode can also be measured in light scattering experiments performed with high numerical aperture objectives (so called ``leakage radiation microscopy'')~\cite{song2011imaging, wang2014controlling, yang2016identification, zhang2018extending, beane2019light}. In contrast, the bound mode does not scatter light into the far field. This means that it is much easier to experimentally study the leaky mode, and for these reasons the discussion below concentrates on the leaky SPP modes.

\begin{figure}[htbp]
\centering
\includegraphics[width=\linewidth]{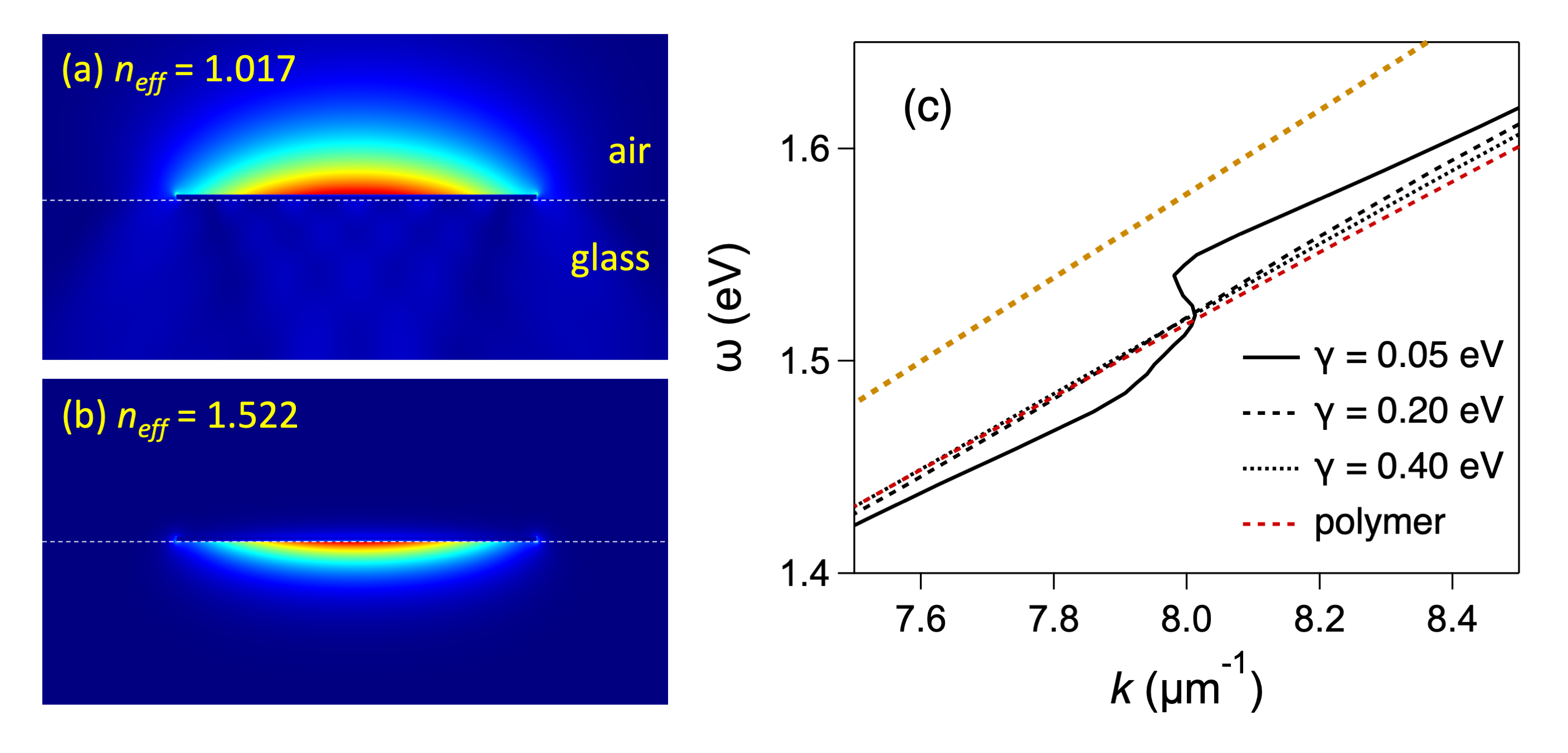}
\caption{(a)--(b) Plots of the magnitude of the electric field ($E/E_{0}$) for the bound and leaky modes for a $4~\mu\text{m}$ wide bare Au nanostripe. The white line shows the interface between the air and glass domains. The intensity scale for the leaky mode is $2\times$ large than that for the bound mode.  The simulations were performed at $\lambda=0.805~\mu m$. (c) Leaky mode dispersion curves for a $4~\mu\text{m}$ wide Au nanostripe with a polymer coating with no dye (``polymer''), and with a 20~nm thick dye layer with $\gamma=0.05, 0.20, 0.40$~eV. The orange dashed line is the light line $\omega=c_{0}k_{0}$.}
\label{fig4}
\end{figure}

Figure~4(c) shows calculated leaky mode dispersion curves for a 20~nm thick dye-polymer layer on top of the nanostripe with values of $\gamma = 0.05, 0.20$ and $0.40$~eV for the electronic transition of the dye. For the $\gamma = 0.20$ and $0.40$~eV simulations the dispersion curves are relatively smooth at frequencies around the dye resonance at 1.5~eV, and very similar to the dispersion curve for a polymer-only coated nanostripe. The group velocity for the leaky mode for these systems is $v_{g} = \partial\omega/\partial k \approx 0.9 c_{0}$.\cite{beane2019light} In contrast, for the $\gamma = 0.05$~eV simulations there is a clear break in the dispersion curve, indicating strong coupling between the leaky SPP mode and the dye transition~\cite{torma2015strong}, and the formation of upper and lower polariton branches. The Rabi frequency, measured as the frequency difference between the extremum in the dispersion curve, is approximately 20~meV for the $\gamma = 0.05$~eV simulations. This is much less than that for the Au nanorod data in Fig.~3(d), and the reason for this difference is discussed below. The calculated dispersion curve for the $\gamma = 0.05$~eV simulations also shows ``back bending''---an apparent negative group velocity---in the avoided crossing region. This has been observed in experiments for molecular layers on metal films~\cite{pockrand1982exciton}, and is a consequence of the damping in the metal and the dye~\cite{pockrand1979exciton}.

Plots of the attenuation versus wavelength are presented in Figs.~5(a)--(c) for nanostripes coated with a polymer layer with no dye, and dye-polymer layers with $\gamma = 0.20$~eV and $\gamma = 0.05$~eV. For the polymer-only coated nanostripes the attenuation is fairly flat in the near-IR region, with approximately equal contributions from radiation damping ($\alpha_{\text{rad}}$) and resistive heating in the Au ($\alpha_{\text{Au}}$).\cite{johns2017dynamics,devkota2019making} At short wavelengths there is an increase in attenuation due to increased radiation damping.\cite{johns2017dynamics} For the $\gamma = 0.20$~eV simulations there is a peak in the attenuation versus wavelength plot that is almost entirely due to absorption in the dye layer ($\alpha_{\text{dye}}$, dashed lines)~\cite{devkota2019making}. The $\gamma = 0.20$~eV simulations correspond to the energy transfer limit discussed above for nanoparticles: the dye causes increased SPP attenuation around the resonance, but the properties of the SPP (its dispersion and group velocity) remain essentially unchanged compared to the system without dye. Note that there is also a slight increase in radiation damping around the dye resonance for $\gamma = 0.20$~eV. This is because the electronic transition of the dye causes a change in the refractive index of the polymer film, in addition to an increase in absorption, see Supporting Information. This changes the field around the nanostructure and, thus, the coupling of the leaky SPP mode to photonic modes in the substrate.\cite{johns2017dynamics,bhandari2025energy,ghosh2025energy}

\begin{figure}[htbp]
\centering
\includegraphics[width=\linewidth]{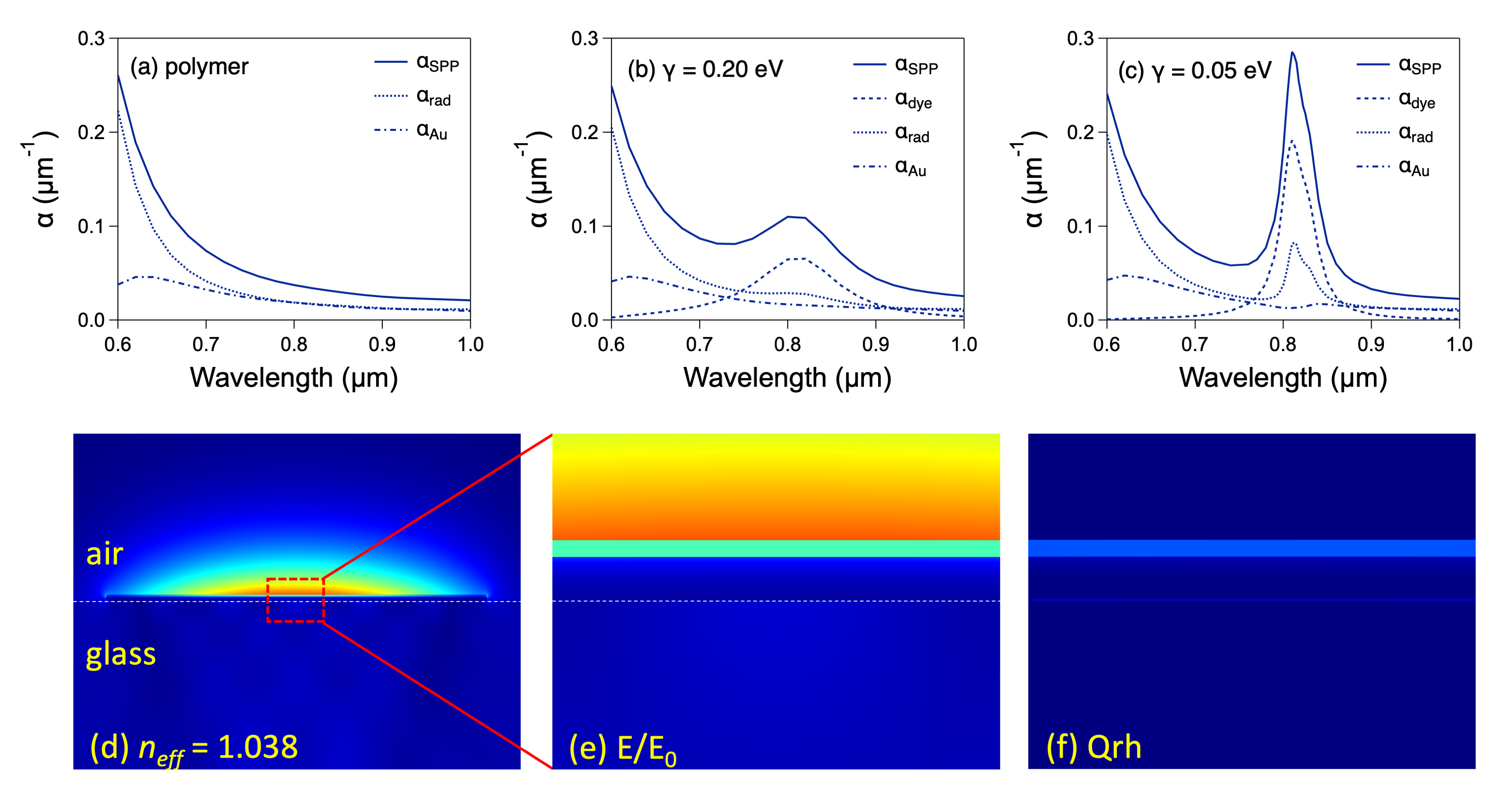}
\caption{(a)--(c) Plots of SPP attenuation for $4~\mu\text{m}$ wide Au nanostripes coated with 20~nm thick polymer layer with no dye, and with a dye layer with $\gamma=0.20$~eV and $\gamma=0.05$~eV. The full lines are the total attenuation, and the dotted, dashed and dash-dotted lines are the attenuation from radiation damping, and resistive heating in the dye and Au layer, respectively. (d) and (e) Plots of the norm of the electric field ($E/E_0$) for $\gamma=0.20$~eV at $\lambda=0.805~\mu\text{m}$ (near the maximum in $\alpha_{\text{SPP}}$). (f) Corresponding plot of the resistive heating $Q_{\text{rh}}$.  At this wavelength there is reasonably strong absorption in the dye layer. Note that the color scales in panels (d)--(f) are the same as those in Fig. 6(d)--(f).}
\label{fig5}
\end{figure}

The physics of the interaction of the dye with the leaky SPP mode for $\gamma = 0.20$~eV is analogous to that for the LSPR in Fig.~2: the field from the SPP mode extends into the dye layer, and absorption in the dye provides an additional dephasing channel that increases the SPP attenuation. The SPP field and the optical losses due to absorption are shown in the plots in Figs.~5(d)--(f). Importantly, for $\gamma = 0.20$~eV the field within the dye layer is smaller than the field in air just outside the nanostructure. Note that it is very hard to see the field in the Au in these plots, as it is strongly attenuated.  Like LSPRs, the attenuation from the dye decreases when a spacer layer is included between the nanostripe and the dye layer. However, because of the different field distribution for the leaky SPP mode compared to the LSPR, the attenuation decreases on a longer length scale than simply the dimensions of the nanostructure.

The plots of SPP attenuation versus wavelength in Fig.~5(c) for $\gamma = 0.05$~eV are similar to that for $\gamma = 0.20$~eV, except that the effects from the dye layer are significantly enhanced. The SPP attenuation due to absorption-induced dephasing ($\alpha_{\text{dye}}$) is much larger, and the increase in radiation damping due to changes in the dielectric environment of the nanostripe ($\alpha_{\text{rad}}$) now has a strong effect on the system. For both the $\gamma = 0.05$~eV and the $\gamma = 0.20$~eV simulations the increase in SPP attenuation traces the dye absorption spectrum, although there is a slight blue shift in the peak that will be discussed in detail below. One of the points made in the discussion of LSPRs given above was that in the strong coupling limit the coupled states share properties. For nanorods at the avoided crossing the linewidths of the $P_{+}/P_{-}$ states are equal, and they both have significant scattering cross-sections (a property of metal particles). For SPPs this behavior can be seen in the dispersion curves. In the avoided crossing region the dispersion curves for the $P_{+}/P_{-}$ states become flatter, which means reduced group velocities. This reflects an averaging of the light-like speeds of SPPs with the stationary excitons. The large attenuation constant at the dye resonance is also a consequence of averaging the SPP and exciton properties: SPPs can have long propagation lengths, whereas, the electronic excitation of the dye does not propagate in our simple model.

The final point considered here is what happens for very sharp dye resonances. Figure~6(a) shows a plot of the real and imaginary components of the dielectric constants for a dye layer with the same parameters as above ($f = 0.1$ and $\omega_{0} = 1.5$~eV), but with $\gamma = 0.03$~eV. This corresponds to a very narrow resonance, but one that is still achievable for J-aggregates at room temperature~\cite{caram2016room, barotov2023near}. For this choice of parameters, the real component of $\epsilon_{\text{dye}}$ is negative in the frequency range around the resonance, see Fig.~6(a)~\cite{pockrand1979exciton}. This can cause quasi-metallic reflection for organic materials, and has been observed for several systems.~\cite{anex1960metallic, hesse1976metallic}. For $\omega_{0}/\gamma \gg 1$ the points where $\text{Re}[\epsilon_{\text{dye}}]$ crosses zero are located at $\omega \approx \omega_{0}$ and $\omega \approx \omega_{0}\sqrt{1+f/\epsilon_{\infty}}$, and give rise to transverse and longitudinal exciton-polariton modes, $\omega_{T}$ and $\omega_{L}$, respectively, in the dye layer~\cite{pockrand1979exciton, hopfield1960some, brillante1978experimental, pockrand1978observation}. This can allow the creation of strongly coupled light-matter states without an optical cavity or plasmonic resonator.\cite{georgiou2020strong, canales2021abundance, pandya2021microcavity} Figure~6(b) shows the dispersion curve obtained from simulations of the coupled Au nanostripe--dye system for $\gamma = 0.03$~eV, with $\omega_{T}$ and $\omega_{L}$ marked by horizontal lines. The system essentially shows two avoided crossings which align with the transverse ($\omega_{T}$) and longitudinal ($\omega_{L}$) exciton-polariton modes of the dye layer.

\begin{figure}[htbp]
\centering
\includegraphics[width=\linewidth]{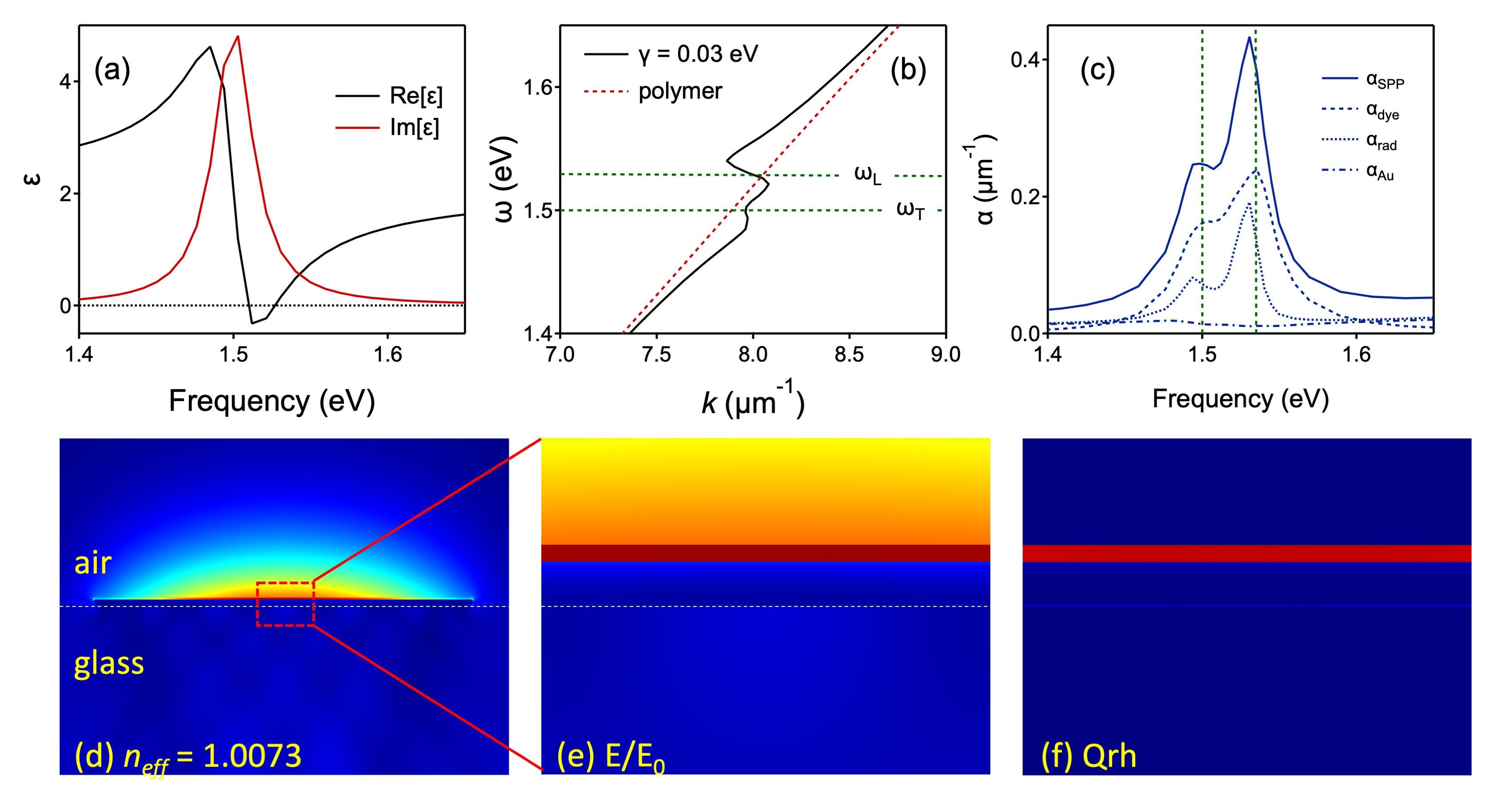}
\caption{(a) Plots of the real (Re$[\epsilon_{dye}(\omega)]$, black line) and imaginary (Im$[\epsilon_{dye}(\omega)]$, red line) components of the dielectric constant for a Drude-Lorentz oscillator with $\gamma=0.03$~eV. (b) and (c) Corresponding dispersion curve and attenuation versus frequency plots for a $4~\mu\text{m}$ wide dye coated Au nanostripe coated with a 20 nm thick dye layer with $\gamma = 0.03$~eV. The green horizontal lines in panel (b) and the vertical lines in panel (c) show the predicted frequencies of the transverse and longitudinal exciton-polariton modes, $\omega_T$ and $\omega_L$, of the dye layer. (d)--(f) Plots of the normalized electric field ($E/E_{0}$) and resistive heating ($Q_{rh}$) for $\gamma=0.03$~eV at $\lambda=0.805~\mu$m (the wavelength corresponding to $\omega_L$). Note that the color scales in panels (d)--(f) are the same as those in Fig. 5(d)--(f)..}
\label{fig6}
\end{figure}

Plots of the attenuation versus frequency for the $\gamma = 0.03$~eV system are presented in Fig.~6(c). The attenuation shows two peaks, one near $\omega_{T}$ and one near $\omega_{L}$. Interestingly, the attenuation is higher at $\omega_{L}$, which is blue shifted from the dye resonance. Figures~6(d) and 6(e) show the SPP mode profiles (norm of the electric field) at $\omega_{L}$ for the dye coated nanostripes. For frequencies around $\omega_{L}$ we now see that the electric field is strongly confined in the dye layer, in contrast to the field plot in Figs.~5(d) and (e) for $\gamma = 0.20$~eV. This causes strong absorption in the dye layer. The field enhancement and increased absorption in the dye layer also happen at $\omega_{T}$, but the effect is weaker. The stronger interaction at $\omega_{L}$ is attributed to a better overlap of the field for the SPP mode (which has significant longitudinal components) with the field from the longitudinal exciton-polariton mode. An important consequence of this is that the attenuation versus wavelength/frequency plot does not trace the dye absorption spectrum. The maximum attenuation is slightly blue shifted from $\omega_{0}$, an effect that can still be seen at larger values of $\gamma$ (see Supporting Information).  

\section{Quantum Model for Polariton Dissipation}
While classical electromagnetic simulations effectively capture the optical response of metal nanostructures, explaining energy dissipation inside strongly coupled light-matter systems requires a quantum electrodynamic framework. In this section, we replace the empirical Drude-Lorentz dielectric layer with a microscopic quantum framework consisting of discrete molecular emitters interacting with quantized localized or propagating plasmonic fields.

We model the coating molecules as an ensemble of $N$ two-level electronic systems with ground state $|{\mathrm{g}_j}\rangle$ and excited state $|{\mathrm{e}_j}\rangle$. All molecular emitters are assumed to have identical transition frequencies $\omega_x = \omega_e - \omega_g$, and pure radiative dephasing rates $\gamma_x$. The transition dipole moment of the $j$-th emitter is $\vec{\mu}_j$. To account for dispersion and absorption of electromagnetic fields induced by metal nanostructures, we employ the macroscopic quantum electrodynamics (MQED) framework, where the quantized electromagnetic environment is captured via the Dyadic Green’s function tensor, $\overline{\overline{\mathbf{G}}}(\vec{r}_i,\vec{r}_j,\omega)$. Physically, the Green’s function captures the electromagnetic response of a molecular dipole at position $\vec{r}_i$ that propagates an electric field to position $\vec{r}_j$ through the complex dielectric environment. Within the MQED framework, the influence of the lossy environment on the molecular ensemble is characterized by the generalized spectral density (GSD) matrix\cite{sanchez-barquilla_few-mode_2022,chuang_tavis-cummings_2022}
\begin{equation}
    J_{\mathrm{rad}(\omega),ij}
    = \frac{\omega^{2}}{\pi \hbar \epsilon_{0}c^2} \vec{\mu}_i\cdot \mathrm{Im}\,\overline{\overline{\mathbf{G}}}(\vec{r}_i,\vec{r}_j,\omega)\cdot \vec{\mu}_j.
    \label{Eq:Gen_Spectral_Density}
\end{equation}
The diagonal elements ($i = j$) correspond to the photonic local density of states (LDOS), which dictates single-molecule spontaneous emission and Purcell factors near the metal surface.\cite{dung_three-dimensional_1998,dung_spontaneous_2000} The off-diagonal elements ($i \neq j$) correspond to the photonic cross density of states (CDOS), describing field-mediated intermolecular interactions.\cite{canaguier-durand_cross_2019, carminati_purcell_2022, ren_quasinormal_2021} The detailed description of the MQED framework is provided in Section S4 of the Supporting Information.

To make quantum dynamics calculations computationally tractable, a wildly adapted approach is to map to the Tavis-Cummings (TC) model. The standard TC model describes the quantized plasmonic field by an effective cavity photon mode of frequency $\omega_c$ with a constant dissipation rate $\gamma_c$, and treats the emitters as coupled oscillators with equal coupling strength $g_c$ to the cavity photon mode.\cite{torma2015strong,chuang_tavis-cummings_2022,sanchez-barquilla_few-mode_2022} In the strong coupling regime, the standard TC Hamiltonian yields two bright polariton states ($|P_\pm\rangle$) along with a manifold of $N-1$ dark states ($\{|D_k\rangle\}$) for $k=1,\cdots,N-1$.  
The upper ($\omega_+$) and lower ($\omega_-$) polariton frequencies can be determined by
\begin{equation}
    \omega_\pm=\frac{\omega_c+\omega_x}{2}\pm\sqrt{\frac{(\omega_c-\omega_x)^2}{4}+Ng_c^2}
\end{equation}
where $\sqrt{N}g_c$ is the collective light–matter coupling strength. 
The upper and lower polaritons are hybrid states formed by mixing the bright collective molecular excitation with the single-photon cavity state. The dark states are purely molecular excitations that have no photonic character and remain uncoupled from the cavity photon.
For this idealized scenario, the decay rates of the $|P_+\rangle$ and $|P_-\rangle$ states due to radiation damping can be estimated by a linear combination of $\gamma_x$ and $\gamma_c$ weighted by their corresponding Hopfield coefficients, $\gamma_+=\cos^2\Theta\gamma_x+\sin^2\Theta\gamma_c$ and $\gamma_-=\sin^2\Theta\gamma_x+\cos^2\Theta\gamma_c$, where the mixing angle is $\Theta=\frac{1}{2}\tan^{-1}\left[\frac{2\sqrt{N}g_c}{\omega_c-\omega_x}\right]$ with $0<\Theta<\frac{\pi}{2}$.

This mapping involves two key simplifications. 
First, all elements of the GSD matrix are assumed to be equal and follow a Lorentzian functional form (see eq. S6 in Supporting Information).\cite{chuang_tavis-cummings_2022} This simplification allows the infinite continuous reservoir of MQED field modes to be mapped to an effective single cavity mode with a damping rate that captures the dissipation into the continuous reservoir. 
Second, the molecular ensemble is assumed to be spatially confined to a region much smaller than the relevant photon wavelength, such that all emitters experience an approximately homogeneous electric field. Within this long-wavelength approximation, the light–matter coupling strength is treated as uniform, as disorder effects arising from molecular vibrational dynamics, solute–solvent collisions, and electromagnetic-field inhomogeneities are neglected.\cite{catuto_interplay_2025,zhou_interplay_2023}
The standard TC model provides an accurate approximation for describing molecular polaritons formed within an optical cavity.\cite{li_molecular_2022} However, in realistic molecule–metal nanostructure systems, these approximations may no longer be valid, thereby introducing additional dissipation channels not captured by the standard TC model, even though the effective energetics of the cavity photon and the resulting polaritons remain well described. In the following sections, we generalize the TC model to incorporate these dissipative processes.

\subsection{Dissipation induced by electromagnetic environments}
We first focus on the dissipation channel induced the radiative environment of LSPRs and SPPs.
To evaluate the matrix element of the GSD as a function of mode frequency $\omega$, we consider two representative emitters within the dye-coating region near a metal nanostructure. The directions of the transition dipole moment are chosen to be in the x-x or z-z dipole configuration, corresponding to J or H aggregates. Using the dielectric constant data for Au from Ref.~\citenum{olmon2012optical}, we calculate the GSD matrix element between the two emitters using eq.~\eqref{Eq:Gen_Spectral_Density}.

\begin{figure}
    \centering
    \includegraphics[width=\linewidth]{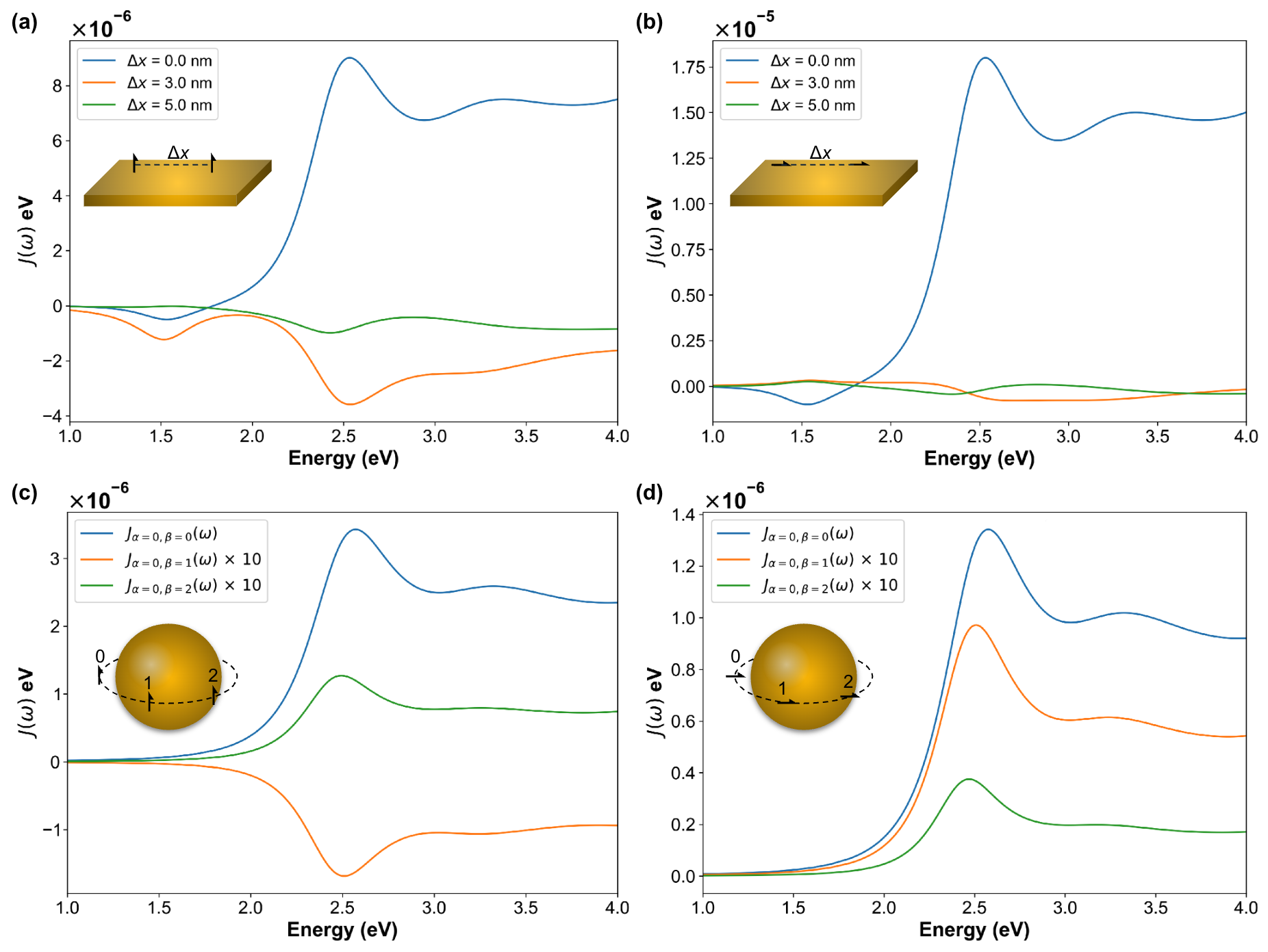}
    \caption{The matrix elements of the GSD $J_\mathrm{rad}(\omega)$ as a function of $\omega$ for representative dipole emitters (a,b) above a Au planar surface at $z=0$ and (c,d) around a Au nanosphere with radius $6~\text{nm}$ centered at the origin. The diagonal element ($\Delta x=0.0~\text{nm}$) and the off-diagonal elements ($\Delta x=3.0,5.0~\text{nm}$) are plotted for the (a) z-z and (b) x-x dipole configurations. For the nanosphere, the emitters are positioned along the equator $\vec{r}_n=(8\cos\frac{2n\pi}{5},8\sin\frac{2n\pi}{5},0)$. The diagonal element ($00$) and the off-diagonal elements ($01,02$) are plotted for the (c) z-z and (d) x-x dipole configurations. Note that the individual matrix elements are not identical and cannot in general be described by a single Lorentzian functional form. }
    \label{figure_7}
\end{figure}

For emitters coupled to SPPs, we consider a planar Au surface of thickness $50\ \text{nm}$ between $z=0$ and $z=-50$ and a glass substrate for $z<-50$. As shown in Fig.~\ref{figure_7}(a--b), two representative emitters are positioned $1~\text{nm}$ above the surface and separated laterally by $\Delta x$. The diagonal term ($\Delta x=0.0~\text{nm}$) exhibits larger values of $J_\mathrm{rad}(\omega)$ than the off-diagonal terms ($\Delta x=3.0,5.0~\text{nm}$). For both the x–x and z–z dipole configurations, the diagonal contribution reaches a maximum near $2.5~\text{eV}$ and shows oscillatory behavior at higher energies. The $2.5~\text{eV}$ threshold corresponds to the onset of the interband transitions in Au,\cite{olmon2012optical,kriebig1995optical,pelton2013introduction,maier2007plasmonics} which enhance decay of the dipole. In addition, for both configurations, the diagonal term presents a negative-valued LDOS feature around $1.5~\text{eV}$, suggesting a net backflow of plasmonic field energy into the emitter. Figure~\ref{figure_7}(a) highlights the z–z dipole configuration where the off-diagonal elements at $\Delta x=3.0, 5.0~\text{nm}$ are mostly negative in the CDOS. This negative CDOS can be attributed to phase differences of the electromagnetic field evaluated at the two emitter positions. Figure~\ref{figure_7}(b) shows the x–x configuration, where the off-diagonal elements ($\Delta x=3.0,5.0~\text{nm}$) are generally much smaller in magnitude than the corresponding diagonal element.

In Fig.~\ref{figure_7}(c--d), we focus on the GSD matrix element for a Au nanosphere of radius $6~\text{nm}$ centered at the origin. One quantum emitter is located at $\vec{r}_0=(8,0,0)$, corresponding to a separation of $2~\text{nm}$ from the sphere surface. A second emitter is positioned at either $\vec{r}_1=(8\cos\frac{2\pi}{5},\,8\sin\frac{2\pi}{5},\,0)$ or $\vec{r}_2=(8\cos\frac{4\pi}{5},\,8\sin\frac{4\pi}{5},\,0)$.
We find that the diagonal element ($00$) also exhibits the largest magnitude, with a pronounced maximum near $2.5~\text{eV}$, and shows oscillatory features at higher energies. We also observe that the overall magnitude of the off-diagonal elements ($01$ and $02$) decreases as the spatial separation increases. Notably, Fig.~\ref{figure_7}(c) shows that the $01$ matrix element becomes negative, whereas the $02$ element is positive. This negative value of the off-diagonal GSD element arises from electromagnetic field interference between two dipoles. In contrast, in Fig.~\ref{figure_7}(d), all matrix elements for the x–x dipole configuration remain positive. These observations indicate that the z–z dipole configuration is sensitive to the electromagnetic phase difference between emitter locations. For the x–x dipole configuration on the equatorial plane, the dipole component along $x$ is insensitive to this phase difference, and therefore preserves a positive sign.

Importantly, Fig.~\ref{figure_7} shows that the individual matrix elements are not identical, and do not follow a Lorentzian functional form. Consequently, for both LSPRs and SPPs, the radiative dissipation cannot be captured within a standard TC model. This finding further suggests that the system’s quantum dynamics should be determined by explicit propagation of the full quantum master equation, incorporating a non-Markovian memory kernel specified by the GSD matrix. However, this does not invalidate the energetics of the polariton and dark states. 


\subsection{Dissipation due to vibrational fluctuations}
Next, we turn our attention to the dissipation channel induced by local nuclear vibrational degrees of freedom in realistic molecular systems. We generalize the TC model to incorporate a Holstein-type harmonic bath composed of independent intramolecular nuclear modes. In this setting, uncorrelated thermal fluctuations of the individual nuclear coordinates break the collective spatial symmetry of the molecular ensemble, thereby relaxing the long-wavelength approximation in the TC model. Within the TC framework, such symmetry breaking opens an intrinsic non-radiative relaxation pathway, enabling bright polaritonic states to scatter into the $N-1$ dark (non-emissive) eigenstates.
To estimate the thermally induced transition rate into the dark-state manifold, we employ second-order perturbation theory together with Fermi’s golden rule, yielding (see Sections S5 of the Supporting Information)
\begin{equation}\label{eq:DS_rate}
     \Gamma_{\pm\rightarrow\mathrm{DS}}=|X|^2\frac{2k_BT}{\hbar}\frac{2\lambda\omega_b}{(\omega_\pm-\omega_x)^2+\omega_b^2}\, .
\end{equation}
Here we adopt a Drude–Lorentz spectral density characterized by the reorganization energy $\lambda$ and bath cutoff (characteristic) frequency $\omega_b$, while $k_BT$ denotes the thermal energy scale of the harmonic bath. 
The factor $|X|^2$ is the Hopfield coefficient quantifying the excitonic content of the polariton: $|X|^2=\cos^2\Theta$ for the upper polariton ($+$) and $|X|^2=\sin^2\Theta$ for the lower polariton ($-$).

\begin{figure}
    \centering
    \includegraphics[width=\linewidth]{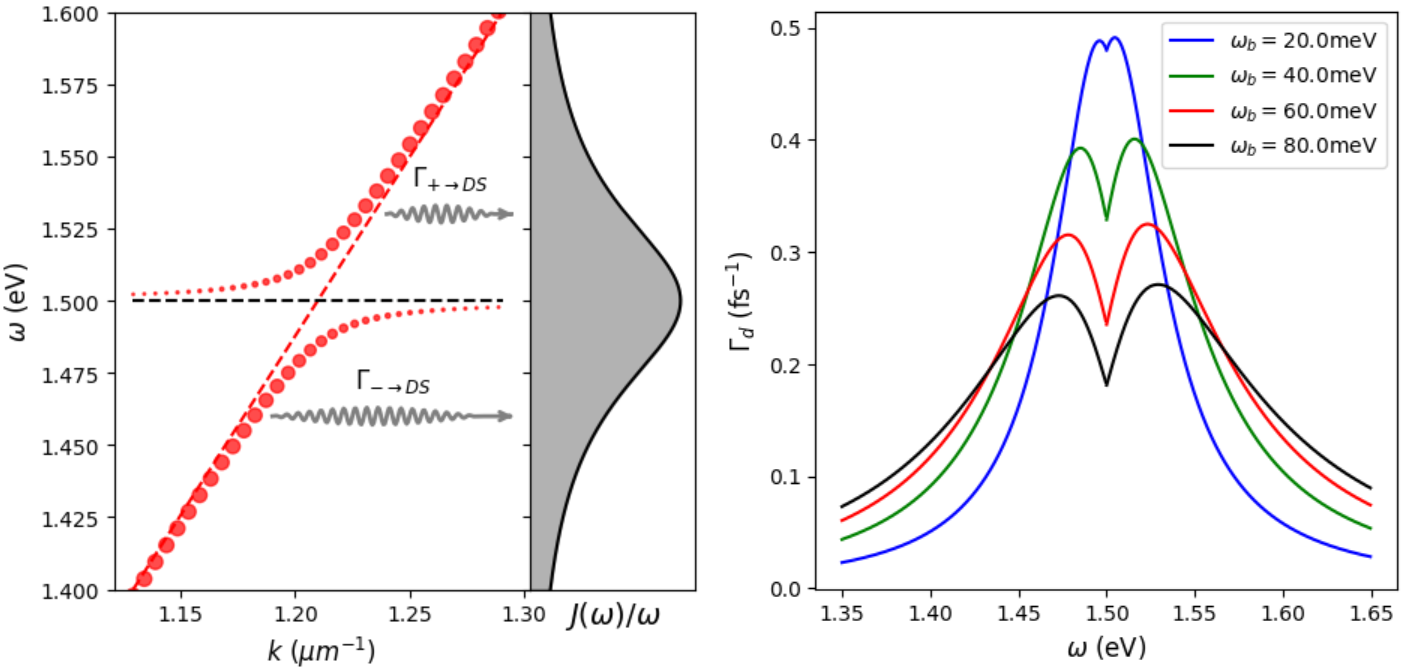}
    \caption{(a) Polariton dispersion curves and their transition to the manifold of dark states (represented by the bath spectral density). The black dashed line indicates the emitter excitation $\omega_x\approx1.5\ \text{eV}$. The red dashed line is the dispersion curve of the plasmonic photon mode. The dots depict the dispersion curves for the upper and lower polaritons with the symbol size representing the Hopfield coefficients of the photon components. 
    (b) The transition rate $\Gamma_{\pm\to\text{DS}}$ as a function of $\omega_c$ for various $\omega_b$. We plot $\Gamma_{+\to\text{DS}}$ for $\omega_c>\omega_x$ and $\Gamma_{-\to\text{DS}}$ for $\omega_c<\omega_x$. Increasing $\omega_b$ shows a clearer dip caused by strong coupling.
    }
    \label{fig:dephasing}
\end{figure}

Figure~\ref{fig:dephasing}(a) illustrates the photon dispersion when coupled to the emitter excitation $\omega_x\approx1.5\ \text{eV}$. We sweep the cavity photon frequency $\omega_c$ and show an avoided crossing at $\omega_c\approx \omega_x$. The collective coupling strength is chosen to be $\sqrt{N}g_c=0.015\ \text{eV}$.
In Fig.~\ref{fig:dephasing}(b), we observe that the transition rate follows the functional form of the vibrational spectral density outside the nearly resonant region ($\omega_c\approx\omega_x$).
In the off-resonant region ($\omega_c\gg\omega_x$ and $\omega_c\ll\omega_x$ ), the Hopfield coefficient becomes independent of $\omega$, i.e  $\cos\Theta=1$ or $\sin\Theta=1$. Thus, the transition rate follows the functional form of the Drude-Lorentz model.
Since optical experiments observe the photon signals with changing $\omega_c$, we plot $\Gamma_{+\rightarrow\mathrm{DS}}$ for $\omega_c > \omega_x$ and $\Gamma_{-\rightarrow\mathrm{DS}}$ for $\omega_c < \omega_x$ in Fig.\ref{fig:dephasing}(b). 

Within the nearly resonant region, the Hopfield coefficients drop from $1$ to $0.5$, manifesting the hybridization between the emitter and photon. At resonance ($\omega_c= \omega_x$), the transition rates of the upper and lower polaritons to the dark states become identical
\begin{equation}
     \Gamma_{+\rightarrow\mathrm{DS}}=\Gamma_{-\rightarrow\mathrm{DS}}=\frac{k_BT}{\hbar}\frac{2\lambda\omega_b}{Ng_c^2+\omega_b^2}
\end{equation}
which is associated with the collective coupling strength $\sqrt{N}g_c$. Consequently, the dip in the transition rate is associated with strong light-matter coupling. We find that, for large $\omega_b$ (fast bath fluctuations), the dip becomes more observable.  

As a final note, we emphasize that, for SPPs coupled to dye molecules, reasonable choices of $\gamma_x$, $\gamma_c$, $\sqrt{N}g_c$, $\omega_b$, and $\lambda$ show that the dissipation rate induced by the nuclear degrees of freedom $\Gamma_{\pm\to\mathrm{DS}}$ is larger than $\gamma_x$ and $\gamma_c$.\cite{bhandari2026plasmon} Consequently, whether estimating $\gamma_\pm$ through the standard TC model or the full GSD matrix, polariton dephasing is dominated by scattering into the vibrational modes of the dark state manifold.

\section{Conclusions and Future Directions}

Coating a metal nanostructure with an absorbing layer creates a composite system that has a complex interaction with light. In this Perspectives article we have used EM simulations to examine how the interaction depends on the strength of the electronic transitions of the absorbing layer. For broad molecular/semiconductor resonances, which have corresponding low peak intensities, the EM simulations show that the metal nanostructure acts as an antenna where the overlap of the field from the nanostructure with the absorbing layer causes increased dephasing of the LSPR/SPP of the metal. For molecular/semiconductor coatings with sharp/intense optical transitions, strong coupling becomes possible. For nanoparticles this splits the LSPR into upper and lower polariton resonances, producing an avoided crossing. At resonance (in the middle of the avoided crossing) the upper and lower polariton states are equal mixtures of the metal and molecular/semiconductor properties: they both show strong absorption and scattering, and their linewidths are equal. For SPPs strong coupling is manifested as a break in the frequency versus wavevector dispersion curve. In this region the SPP is strongly attenuated by absorption in the dye/semiconductor layer. For molecular/semiconductor systems with very strong absorption, where the real part of the dielectric constant of the absorbing layer goes negative, the structure of the simulated SPP dispersion curves becomes complicated, with multiple avoided crossings. This is attributed to the presence of exciton-polariton modes in the absorbing layer, as has been observed in experiments with Ag films~\cite{pockrand1982exciton}.

The interaction between molecular/semiconductor systems and plasmons can also be viewed through the lens of the MQED framework, which describe hybridization between quantum emitters and quantized electromagnetic fields of metal nanostructures. We emphasize that in general the standard Tavis–Cummings (TC) model is not sufficient for quantitatively describing energy dissipation of the hybrid system. In particular, for radiative dissipation, the GSD matrix elements are not equal around the onset of the interband transitions and do not follow a Lorentzian form, which violates the TC model’s assumption of a constant dissipation rate for the cavity photon mode. This suggests that an explicit treatment of the full GSD matrix is required to accurately compute the hybrid-system dynamics.  The effect of nuclear vibrations can be included in the TC framework through incorporating Holstein-type vibrational baths, which provide relaxation pathways that cause the polaritonic states to decay into the dark-state manifold.\cite{bhandari2025energy} This generalization enables the description of additional dephasing and relaxation channels beyond those captured by the simple rate estimates in the TC model.  Comparison of $\Gamma_{\pm\rightarrow\mathrm{DS}} $ to estimates of the radiative decay from the coupled oscillator model show that scattering into dark states should dominate the dephasing of the polariton states.  This picture agrees with the EM simulation results for coupled SPP-exciton states, which show that main contribution to attenuation is resistive heating in the dye.

Although the basic physics is the same for LSPRs and SPPs coupling to molecular or semiconductor excitations, there are several important differences for designing experimental studies. First, in general the Rabi frequencies for LSPRs are larger than those for SPPs. This is an advantage for using LSPRs to study strong coupling effects. A disadvantage is that creating a dispersion curve is much more challenging for LSPRs. In Figure~3(d) each point along the frequency versus aspect ratio plot represents a different nanoparticle.\cite{oh2026plasmon,zengin2013approaching} In contrast, for the leaky SPP mode, dispersion curves and attenuation versus wavelength plots can be obtained from single nanostructures by simply changing the excitation wavelength. This is much more straightforward and controlled, and directly yields data that can be compared to the MQED calculations (Fig. 8(b)).  Finally, the way the interaction between the plasmonic structure and the molecular/semiconductor layer changes with distance is quite different. In both cases the interaction depends on the overlap of the plasmonic field with the molecular/semiconductor layer. For LSPRs the field decays on the length scale of the dimensions of the structure~\cite{maier2007plasmonics, pelton2013introduction}, whereas for SPPs the field is less confined, so the decay occurs over a longer length scale~\cite{johns2017dynamics}.   

We envision several important future directions.  From the MQED side, development of a full quantum master equation approach is needed to account for the complex structure of the GSD matrix in the presence of interband transitions in the metal.  From the experimental side, detailed measurements are needed for strongly coupled SPP-exciton states to map the attenuation versus wavelength.  Both the MQED calculations and the EM simulations predict structure around the avoided crossing region, and it will be interesting to see if this can be observed.

\section*{Supporting Information}
Refractive index and dielectric constant data for the model dye layer at different values of $\gamma$; Plots of LSPR spectra for $\gamma=0.03$~eV and different aspect ratios; Comparison of SPP attenuation and dye absorption versus wavelength for different values of $\gamma$; From macroscopic quantum electrodynamics (MQED) framework to Tavis-Cummings model; Derivation of the dissipation rate induced by Holstein-type vibrational coupling.

\section*{Acknowledgements}
J.B., A.S. and G.V.H. acknowledge support from the National Science Foundation through grant CHE-2304905, and G.L. and H.-T.C acknowledge support from the National Science Foundation through grant CHE-2542091.

\bibliography{references}
\end{document}